\documentclass[preprint]{aastex}
\usepackage{natbib}
\usepackage{hyperref}
\usepackage{rotating}
\usepackage{graphicx}
\usepackage{lscape}
\usepackage{float}
\newcommand{\kms}{km\,s$^{-1}$}
\newcommand{\mdot}{\ensuremath{\dot{M}}}              
\newcommand{\tauross}{\ensuremath{\tau_{\mathrm{ross}}}}     
\newcommand{\vwind}{\ensuremath{\mathit{v}_\mathrm{wind}}}   
\newcommand{\msunyr}{\ensuremath{{\rm M}_{\odot}\,{\rm yr}^{-1}}}  
\newcommand{\lsun}{\ensuremath{{\rm L}_{\odot}}} 

\shorttitle{SN~1998S: Early Emission in High Resolution}
\shortauthors{Shivvers et al.}

\begin{document}
\title{Early Emission from the Type IIn Supernova 1998S at High Resolution}
\author{Isaac Shivvers\altaffilmark{1}, 
              Jose H. Groh\altaffilmark{2}, 
              Jon C. Mauerhan\altaffilmark{1}, 
              Ori D. Fox\altaffilmark{1},
              Douglas C. Leonard\altaffilmark{3}, and
              Alexei V. Filippenko\altaffilmark{1}}
\email{ishivvers@astro.berkeley.edu}
\altaffiltext{1}{Department of Astronomy, University of California,
    Berkeley, CA 94720-3411}
\altaffiltext{2}{Geneva Observatory, Geneva University,
    Chemin des Maillettes 51, CH-1290 Sauverny, Switzerland}
\altaffiltext{3}{Department of Astronomy, San Diego State University,
    San Diego, CA 92182}

\begin{abstract}
The well-studied Type IIn supernova (SN) 1998S is often dubbed the prototypical SN IIn, and it provides a unique opportunity to study its progenitor star from within as the SN lights up dense circumstellar material (CSM) launched from the progenitor. Here we present a Keck HIRES spectrum of SN 1998S taken within a few days after core collapse --- both the earliest high-resolution ($\Delta\lambda < 1.0$\,\AA) spectrum published of a SN IIn and the earliest published spectrum of SN 1998S. 
Modern SN studies achieve impressively short turn-around times between SN detection and the first observed spectrum, but high-resolution spectra of very young supernovae are rare; the unique spectrum presented here provides a useful case study for observations of other young SN systems including SN~2013cu, which displayed a remarkably similar spectrum when very young. We examine the fully resolved emission-line profiles of SN~1998S, finding evidence for extreme mass loss from the progenitor at velocities much less than those characteristic of Wolf-Rayet (WR) stars. We model our high-resolution SN~1998S spectrum using the radiative-transfer code CMFGEN and explore the composition, density, and velocity gradients within the SN system.  We find a mass-loss rate of $6.0 \times 10^{-3}$\,\msunyr\ during the $\sim 15$\,yr before core collapse, while other studies indicate a much lower rate at earlier times  ($>$15\,yr before core collapse).
A comparison with a spectrum of SN~2013cu indicates many similarities, though SN~2013cu was of Type IIb --- indicating that very different supernovae can arise from progenitors
with extreme mass loss in the last few years before explosion.
\end{abstract}


\section{Introduction}

Type IIn supernovae (SNe) provide unique observational windows into the end stages of evolution in massive stars.
Though they are somewhat rare \citep[making up $\sim 9$\,\% of all core-collapse SNe;][]{li11,smith11},
SNe~IIn display a large diversity in observed properties. 
The class is identified by the presence of narrow or intermediate-width emission lines
(full width at half-maximum intensity [FWHM] $\lesssim 1000$\,\kms),  created by radiative recombination and electron scattering 
in relatively slow-moving circumstellar material (CSM) exterior to the expanding ejecta \citep[e.g.,][]{schlegel90,filippenko91,chugai91,filippenko97,fransson02}.  
The presence of this CSM around SNe~IIn strongly affects many of their observational traits \citep[e.g.,][]{chugai02,chevalier11,fransson14}. To give just a few examples,
the development of a dense shell of swept-up material in the post-shock medium can suppress or mask the underlying broad
ejecta features typically seen in normal SN spectra; if the CSM is optically thick, the light-curve shape can be altered
and the radiation of the developing SN must diffuse outward through the CSM; 
and the ejecta-wind interaction can contribute luminosity to the SN (especially at late times).

The commonly accepted explanation for the relatively dense CSM around SNe~IIn requires 
significant mass loss from the progenitor stars in the years and decades before they undergo core collapse.
A broad range in mass-loss rates from a diversity of possible progenitors yields a wide variety in observed properties \citep[e.g.,][]{kiewe12}. 
SNe~IIn can be some of the most luminous SNe in the Universe, efficiently converting the kinetic energy of the expanding SN ejecta into light through shock interaction
with the CSM \citep[e.g.,][]{ofek07,smith10}, and they appear to create large amounts of dust compared to other
SN classes \citep[e.g.,][]{fox11}.

SN~1998S was discovered in NGC~3877 on 1998 March 2.68 
(UTC dates and times are used throughout this article),
with a nondetection $\sim 3$\,d earlier
\citep[][]{li98,qiu98}.  This indicates that SN~1998S was discovered within 
a few days of core collapse \citep[][]{leonard00}, though the exact age of the SN is uncertain. Accordingly, all epochs used here 
are relative to the discovery date. 
SN~1998S is one of the best-studied examples of the SN~IIn class, 
with well-sampled optical and infrared light curves \citep[e.g.,][]{fassia00,gerardy02},
low-resolution optical spectroscopic and spectropolarimetric observations \citep[e.g.,][]{leonard00,anupama01,wang01}, very late-time
spectroscopic observations \citep[e.g.,][]{mauerhan12}, space-based UV spectroscopy \citep{fransson05}, X-ray and
radio observations \citep{pooley02}, and high-resolution optical spectroscopy \citep[e.g.,][]{bowen00,fassia01},
as well as modeling efforts \citep[e.g.,][]{chugai01,lentz01,chugai02}.

In this paper, we present a single epoch of high-resolution echelle spectroscopy of SN~1998S observed on 1998 March 4 --- 
the earliest spectrum of this object yet published (1.86\,d after discovery),
and (we believe) the earliest high-resolution ($\Delta \lambda < 1.0$\,\AA) spectrum of a SN~IIn ever taken. 
We also perform a detailed radiative-transfer model using the CMFGEN code \citep{hm98},
following the methodology described by \citet{groh14}. This allows us to explore the temperature, density, composition,
and ionization structure of the SN system based on the high resolution spectrum.

Capturing a spectrum of a SN while it is very young is a serious challenge, but recent technical advances have 
begun to push the envelope for what are called ``early'' spectra of core-collapse SNe --- 
in some cases perhaps even probing the fading emission attributed to shock breakout from the progenitor
star \citep[``flash spectroscopy'';][]{agy}. Though there are significant differences 
between the physical environment of SN~1998S and the environs of other SNe not of Type IIn, there are also many similarities, especially
when considering the spectral signatures of progenitor winds illuminated by shock breakout from the star or by ongoing shock interaction with a dense CSM.
We believe that the spectrum and model given here can serve as a useful case study when interpreting early-time spectra of many core-collapse SNe.
In \S\,\ref{sec:obs} we discuss our data and reduction techniques, in \S\,\ref{sec:model} we describe our modeling techniques and
the best-fit CMFGEN model of the young SN~1998S system, while \S\,\ref{sec:discussion} synthesizes
these results with other studies of SN~1998S and additional objects.
We summarize and conclude in \S\,\ref{sec:conclusion}.

\section{Observations and Data Reduction}
\label{sec:obs}

Between 12:45 and 13:20 on 1998 March 04 (UT),
we obtained a set of spectra of SN~1998S 
with the Keck High-Resolution Echelle Spectrometer \citep[HIRES;][]{vogt94}, exposing
for $2 \times 500$\,s in each of two setups, {\em h03} and {\em h07}.  Setup {\em h03} covers a
wavelength range of 3875--6270\,\AA\ with a measured resolution of $\sim 0.3$\,\AA, while setup {\em h07} covers 5065--7480\,\AA\ at
a resolution of $\sim 0.2$\,\AA.  (Quoted resolutions are averaged FWHM measures from Gaussian fits to narrow Milky Way \ion{Na}{1}\,D
lines and telluric absorption lines in our reduced spectra.)

At this time, SN~1998S exhibited a visual magnitude of $\sim 13.5$ \citep{li98}.
Photometry of NGC~3877 at the location of the SN, performed with {\em g}-band images from the Sloan Digital Sky Survey,
indicates that the explosion site has a surface brightness of $\sim 20$--21\,mag\,arcsec$^{-2}$. 
Our observations were obtained with a slit width of $\sim 1.0$\,arcsec, comparable to the seeing, and so any contamination of the SN spectrum 
by host-galaxy light is quite small.
For each setup the two 500\,s exposures were reduced using the MAKEE
program\footnote{\href{http://www2.keck.hawaii.edu/inst/common/makeewww/}{www2.keck.hawaii.edu/inst/common/makeewww}}
and coadded. Unfortunately, obtaining an accurate flux normalization and calibration is quite difficult
for echelle spectrographs like HIRES.
However, because the continuum of SN~1998S is remarkably smooth and well-defined 
during the first several days \citep{leonard00,fassia01},
we are able to remove the convolved SN continuum and flatfield imperfections by fitting
for and dividing by a smooth pseudo-continuum both across orders and within each individual order
of our HIRES spectra. 

We first fit a smooth nonparametric function to the median values of each order to determine the 
response function along the $y$ axis of the CCD (the axis along which echelle orders are spread).
To account for the few prominent broad-wavelength features in our spectra, we do not include the values
from orders heavily affected by these emission features.
After dividing our spectrum by the CCD's $y$-axis response function, we fit another smooth nonparametric
function within each individual order, and again divide our spectrum by it, thereby accounting for
the variations of the SN continuum and flatfield along the $x$ axis of the CCD (the axis along which each
individual order is dispersed).  As above, we must be concerned
about the few broad features, so we interpolate the response functions of neighboring
orders onto those orders with known broad emission lines.  The $x$-axis response functions change slowly
with order number and this procedure appears to approximate the true response function well.
The \ion{C}{3}\,$\lambda 5696$ emission line was clearly detected in both setups {\em h03} and {\em h07} and provides a quick check that our
wavelength and relative-flux procedures produce a well-calibrated spectrum: fitting a simple Gaussian to the line
in each setup yields line centers that match to within 0.05\,\AA~and amplitudes that match to within 0.2\%.
All narrow features and the major broad emission features are preserved with this process, but 
note that subtle broad features are very difficult to disentangle from the individual echelle order responses
and are therefore likely to be subtracted out of our spectrum.

The result of the above procedure is a spectrum in units of $F_{\lambda} / F_{\lambda,c}$, where
$F_{\lambda,c}$ is the flux of the SN continuum per unit wavelength.
To get a spectrum in 
units of $F_{\lambda}$ we must determine $F_{\lambda,c}$.
We approximate $F_{\lambda,c}$ for the March 4 spectrum by measuring it from 
a well-calibrated Keck Low-Resolution Imaging Spectrometer (LRIS) observation taken on March 6 \citep{leonard00}.
At these early times the SN continuum is smooth, there are no broad absorption features, and the
SN exhibits an ultraviolet (UV) excess compared to a simple blackbody; we choose to define the continuum
with a smooth nonparametric spline function. We assume that the continuum shape is changing only slowly at early times 
\citep[as implied by spectra from days 3, 4, and 5;][]{leonard00}, and that the continuum shape
on day 2 is similar to that shown on day 4. Regardless, the largest source of uncertainty
in this relative flux calibration is attributable to the unknown color evolution of SN~1998S
between days 2 and 4.

Though the color evolution at these times was small, the total-flux evolution was not.  
SN~1998S was rapidly increasing in brightness over this timespan, so we perform an 
additional absolute-flux correction by matching the spectrum to the unfiltered
photometry observed by the Katzman Automatic Imaging Telescope \citep[KAIT;][]{filippenko01}
the same night \citep[$\sim 13.5$\,mag;][]{li98}.
Though the uncertainty on our absolute-flux calibration procedure is large,
this photometric data point falls smoothly between the unfiltered magnitude from the discovery image
a little more than one day before \citep[$\sim 15.2$\,mag;][]{qiu98} and the later, more
densely sampled unfiltered and filtered observations \citep[e.g.,][]{fassia00,anupama01}.
We estimate that the uncertainty of the calibrated absolute-flux level is $\sim 20$\%.

\begin{figure}[h]
\begin{centering}
\includegraphics[width=0.75\textwidth]{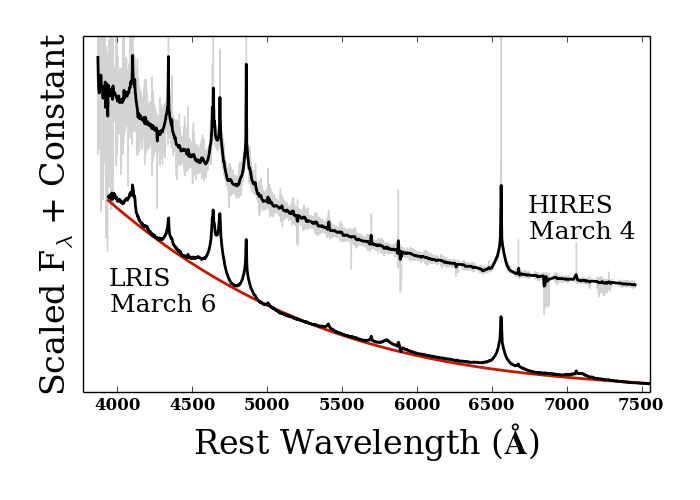}
\caption{Our approximately flux-calibrated HIRES spectrum of SN~1998S observed 1998 March 4 (day 2)
         and a well-calibrated LRIS spectrum taken 2\,d later (day 4); both are shown before
         application of any dust reddening correction.  The full HIRES spectrum is shown
         in gray, with a spectrum binned to the approximate resolution of the LRIS spectrum
         overlain in black.  The LRIS spectrum is shown at bottom along with the measured
         SN continuum (red), which has been applied to the HIRES spectrum during the approximate
         flux-normalization procedure (see \S\ref{sec:obs}). Note that our HIRES flux calibration
         procedure may have removed weak and broad features present in the March 4th spectrum ---
         e.g. the wings of the weaker helium lines or the features near 5801 and 7115\,\AA~visible on March 6.
          \label{fig:fluxnorm}}
\end{centering}
\end{figure}

Regardless of the significant uncertainties in the flux calibrations,
the widths, positions, and shapes of all emission lines are well determined.
Figure~\ref{fig:fluxnorm} shows our final HIRES spectrum alongside the well-calibrated LRIS spectrum
for comparison.
Note that our spectral coverage is not complete over the full 3875--7480\,\AA\ range;
there are gaps between many of the spectral orders
(complete coverage at the blue end spreads out to a $\sim 50$\,\AA\ gap between the two reddest orders).
Our final composite spectrum includes coverage from setup {\em h03}
over the range 3875--6265\,\AA\ and from setup {\em h07} over the 
range 6275--7480\,\AA; upon publication it will be made available in electronic format on WISeREP
\citep[the Weizmann Interactive Supernova data REPository;][]{wiserep}.\footnote{\url{http://www.weizmann.ac.il/astrophysics/wiserep .}}

Previous authors present total interstellar reddening values toward SN~1998S of
$E(B-V) \approx 0.23 \pm 0.1$\,mag \citep{leonard00,fassia00}, but recent studies have re-evaluated and improved upon
the procedures used to calculate that value. Using the scaling relations published by 
\citet{poznanski12} and equivalent widths measured from our flux-normalized HIRES spectrum 
(both components of the host galaxy's Na D line are well resolved), we calculate that
$E(B-V)_{\rm host} = 0.128^{+0.040}_{-0.028}$\,mag.  Including the \citet{schlafly11} measurement
of $E(B-V)_{\rm MW} = 0.0202 \pm 0.0009$\,mag, we derive that the total interstellar reddening
toward SN~1998S is $E(B-V) = 0.148^{+0.040}_{-0.028}$\,mag.
As described in greater detail in \S\ref{sec:model}, we adopt a redshift of $z = 0.00286$.
Note that this is slightly at odds with the best-fit value of $z = 0.002824 \pm 0.0000097$ 
presented by \citet[][]{fassia01}, and is slightly different from the measured value for the host galaxy
NGC~3877 \citep[$0.002987 \pm 0.000013$;][]{verheijen01}.
We apply appropriate redshift and reddening corrections to all spectra before analysis, assuming a \citet{cardelli89}
dust law and $R_{V} = 3.1$. However, we allow the host-galaxy reddening to vary when fitting a model of the system to the spectrum;
see the extended extinction discussion in \S\ref{sec:reddening}.

\subsection{Young SN~1998S in High Resolution}
\label{sec:analysis}

\begin{landscape}
\begin{figure*}
\vspace*{-2cm}
\hspace*{-2cm}
\includegraphics[width=1.65\textwidth]{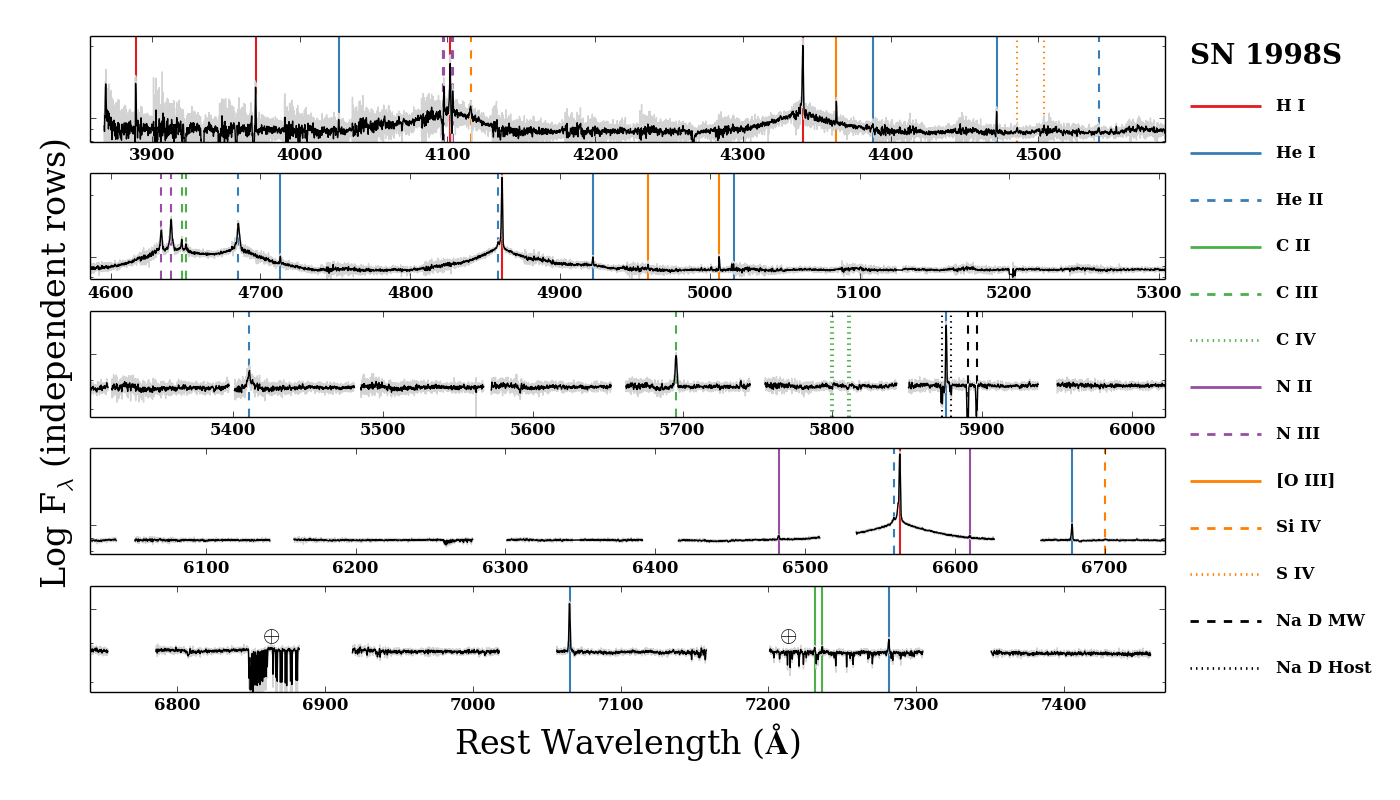}
\caption{A HIRES spectrum of SN~1998S taken on 1998 March 4 (day 2), with all identified
                lines marked.  Full spectrum shown in gray, overlain by a spectrum smoothed
                to a resolution of $\sim 0.5$\,\AA~in black.
                A continuum fit has been subtracted and each row has been individually scaled along
                the ordinate axis for clarity.
                \label{fig:fullpagespec}}
\end{figure*}
\end{landscape}

Figure \ref{fig:fullpagespec} shows our 1998~March~4 HIRES spectrum in high resolution with all identified emission
lines labeled. All line identifications have been confirmed with the help of the CMFGEN model.
Uncorrected telluric absorption features are apparent in our spectrum; we
make sure they do not overlap with the features of interest and otherwise ignore them.  For each identified line
we either fit a single Gaussian profile or, for the strongest few Balmer lines which clearly display prominent extended
wings, the sum of a Gaussian profile and a modified Lorentzian profile (where the exponent is allowed to 
deviate from 2.0). For each line we perform a maximum-likelihood fit and estimate our parameter errors using
Markov-Chain Monte Carlo methods; the results are listed in Table \ref{tab:lines}.

\begin{deluxetable}{l c c c c}
\tablecolumns{5}
\tablecaption{Detected Spectral Lines\label{tab:lines}}
\tablewidth{0pt}
\tablehead{
\colhead{Ion} & \colhead{Rest wavelength} & \colhead{Observed wavelength$\pm$1$\sigma$} &
                \colhead{Doppler velocity$\pm$1$\sigma$} & \colhead{FWHM$\pm$1$\sigma$} \\
\colhead{} & \colhead{(\AA)} & \colhead{(\AA)} & \colhead{(\kms)} & \colhead{(\kms)}
}
\startdata
H I & 3889.06 & 3889.03$\pm$0.035 & -2.2$\pm$2.70 & 29.76$\pm$2.783 \\ 
H I & 3970.08 & 3970.16$\pm$0.122 & 5.8$\pm$9.19 & 60.18$\pm$7.853 \\ 
He I & 4026.19 & 4026.43$\pm$0.017 & 17.4$\pm$1.29 & 29.93$\pm$3.291 \\ 
N III$^\ddag$ & 4097.36 & 4097.59 & 16.7 & 31.97 \\ 
N III$^{\ddag*}$ & 4097.36 & 4096.95 & -29.9 & 78.2 \\ 
H I & 4101.73 & 4101.82$\pm$0.013 & 6.2$\pm$0.91 & 43.93$\pm$1.031 \\ 
N III$^\ddag$ & 4103.39 & 4103.63 & 17.7 & 39.31 \\ 
N III$^{\ddag*}$ & 4103.39 & 4102.94 & -32.8 & 79.4 \\ 
Si IV & 4116.10 & 4115.69$\pm$0.130 & -30.2$\pm$9.50 & 102.70$\pm$15.827 \\ 
H I & 4340.47 & 4340.55$\pm$0.004 & 5.2$\pm$0.29 & 52.63$\pm$7.542 \\ 
H I$^\dag$ & 4340.47 & 4338.73$\pm$0.371 & -120.4$\pm$25.62 & 3533.02$\pm$36.013 \\ 
$[$O III$]$ & 4363.21 & 4363.28$\pm$0.044 & 4.9$\pm$3.03 & 33.60$\pm$4.356 \\ 
He I$^\ddag$ & 4388.15 & 4387.88 & -18.4 & 72.49 \\ 
He I & 4471.48 & 4471.72$\pm$0.006 & 16.4$\pm$0.43 & 29.90$\pm$0.898 \\ 
S IV & 4485.64 & 4485.61$\pm$0.055 & -2.2$\pm$3.65 & 49.86$\pm$7.024 \\ 
S IV & 4504.11 & 4503.86$\pm$0.084 & -16.6$\pm$5.57 & 52.98$\pm$11.734 \\ 
He II & 4541.49 & 4540.86$\pm$0.044 & -41.3$\pm$2.88 & 75.32$\pm$10.595 \\ 
N III & 4634.14 & 4633.86$\pm$0.045 & -18.1$\pm$2.92 & 76.98$\pm$9.911 \\ 
N III & 4640.64 & 4640.41$\pm$0.107 & -14.8$\pm$6.93 & 90.31$\pm$8.120 \\ 
C III & 4647.42 & 4647.52$\pm$0.015 & 6.7$\pm$0.98 & 40.45$\pm$2.084 \\ 
C III & 4650.25 & 4650.35$\pm$0.057 & 6.4$\pm$3.69 & 38.23$\pm$5.538 \\ 
He II & 4685.80 & 4685.35$\pm$0.015 & -28.7$\pm$0.98 & 122.71$\pm$3.173 \\ 
He I & 4713.15 & 4713.30$\pm$0.037 & 9.7$\pm$2.38 & 34.73$\pm$3.931 \\ 
He II$^\ddag$ & 4859.32 & 4858.72 & -36.9 & 50.16 \\ 
H I & 4861.35 & 4861.32$\pm$0.002 & -1.9$\pm$0.10 & 43.72$\pm$3.626 \\ 
H I$^\dag$ & 4861.35 & 4860.44$\pm$0.026 & -56.0$\pm$1.58 & 2241.28$\pm$62.489 \\ 
He I & 4921.93 & 4922.00$\pm$0.016 & 4.5$\pm$0.98 & 36.30$\pm$1.535 \\ 
$[$O III$]$ & 4958.91 & 4958.85$\pm$0.031 & -3.7$\pm$1.86 & 29.62$\pm$4.329 \\ 
$[$O III$]$ & 5006.84 & 5006.33$\pm$0.052 & -30.9$\pm$3.10 & 32.51$\pm$4.197 \\ 
He I$^\ddag$ & 5015.68 & 5015.90 & 13.2 & 20.32 \\ 
He II & 5411.52 & 5410.65$\pm$0.061 & -48.0$\pm$3.40 & 116.45$\pm$9.634 \\ 
C III & 5695.92 & 5695.69$\pm$0.021 & -11.9$\pm$1.13 & 73.26$\pm$1.353 \\ 
C IV$^{\ddag*}$ & 5801.31 & 5799.03 & -117.6 & 125.3 \\ 
C IV$^{\ddag*}$ & 5811.97 & 5810.25 & -88.6 & 86.9 \\ 
He I & 5875.62 & 5876.01$\pm$0.082 & 19.9$\pm$4.16 & 39.44$\pm$5.623 \\ 
N II & 6482.05 & 6482.37$\pm$0.083 & 14.8$\pm$3.84 & 43.15$\pm$2.900 \\ 
He II$^\ddag$ & 6560.09 & 6559.25 & -38.2 & 44.56 \\ 
H I & 6562.82 & 6563.18$\pm$0.001 & 16.4$\pm$0.06 & 42.57$\pm$3.467 \\ 
H I$^\dag$ & 6562.82 & 6562.21$\pm$0.023 & -27.7$\pm$1.03 & 1388.78$\pm$13.718 \\ 
N II & 6610.56 & 6609.89$\pm$0.046 & -30.3$\pm$2.07 & 26.76$\pm$4.336 \\ 
He I & 6678.15 & 6678.19$\pm$0.007 & 1.8$\pm$0.34 & 36.32$\pm$0.584 \\ 
Si IV & 6701.21 & 6700.54$\pm$0.102 & -30.0$\pm$4.54 & 75.74$\pm$13.493 \\ 
He I & 7065.19 & 7065.46$\pm$0.045 & 11.3$\pm$1.93 & 41.07$\pm$2.758 \\ 
C II & 7231.32 & 7231.31$\pm$0.044 & -0.4$\pm$1.82 & 37.98$\pm$5.398 \\ 
C II & 7236.42 & 7236.42$\pm$0.036 & -0.2$\pm$1.50 & 25.73$\pm$2.299 \\ 
He I & 7281.35 & 7281.48$\pm$0.076 & 5.4$\pm$3.15 & 33.64$\pm$4.743 \\ 
\enddata
\tablecomments{Identified absorption and emission lines, as well as measures of the velocity offset
                              and full width at half-maximum intensity (or half-minimum intensity, for absorption lines)
                              as determined through a maximum-likelihood fit.  Lines were fit with a Gaussian profile unless otherwise noted,
                              and all errors are 1$\sigma$ confidence levels estimated by Markov-Chain Monte-Carlo sampling, but
                              do not include any systematic errors due, for example, to biases imparted by the instrument or reduction method. 
                              Note that our error estimation method fails 
                              for lines with confounding features very nearby.
                              Line identifications made with the assistance of our CMFGEN model and the NIST Atomic Spectral Database$^{**}$.}
\tablenotetext{\dag}{\,Broad line fit by a modified Lorentzian profile with a variable exponent.}
\tablenotetext{\ddag}{\,MCMC error analysis failed or not applicable.}
\tablenotetext{*}{\,Absorption feature.}
\tablenotetext{**}{\,\href{http://www.nist.gov/pml/data/asd.cfm .}{www.nist.gov/pml/data/asd.cfm .}}
\end{deluxetable}

\section{Modeling}
\label{sec:model}

\subsection{CMFGEN}
\label{sec:cmfgen}

We use the radiative-transfer code CMFGEN \citep{hm98} to analyze the high-resolution spectrum of SN~1998S. The setup of the code for a SN interacting with its CSM follows that of \citet{groh14}, so here we only briefly recall the main features. CMFGEN computes the radiative transfer in spherical symmetry, simultaneously solving for the level populations and radiation field under nonlocal thermodynamic equilibrium conditions. The effects of line blanketing are taken into account following a super-level approach and, for each element in the model, the most significant species are included.

The properties of our model are determined by the radius of the inner boundary ($R_\mathrm{in}$), bolometric luminosity ($L_\mathrm{SN}$), and the chemical abundances of the elements listed in Table~\ref{tab:model}. Since no hydrodynamical modeling is performed, we assume a constant velocity (\vwind) and mass-loss rate (\mdot) for the progenitor wind. The wind density structure is determined according to the equation of mass continuity, assuming $\rho \propto r^{-2}$. At the inner boundary, we assume a  steep density gradient with a scale height of $0.007\,R_\mathrm{in}$, as theoretically expected for interacting SNe \citep[e.g.,][]{chugai01,chevalier11}.  This ensures that  the inner boundary has high optical depths (Rosseland optical depth $\tauross = 50$ at $R_\mathrm{in}$) and that the diffusion approximation holds at the inner boundary. Our models also assume that all energy is generated at distances $r<R_\mathrm{in}$, that time-dependent effects are negligible, and that the medium is unclumped.

\subsection{A Model of SN~1998S in High Resolution}
\label{sec:98s_model}

\begin{figure}[H]
\begin{centering}
\includegraphics[width=0.99\textwidth]{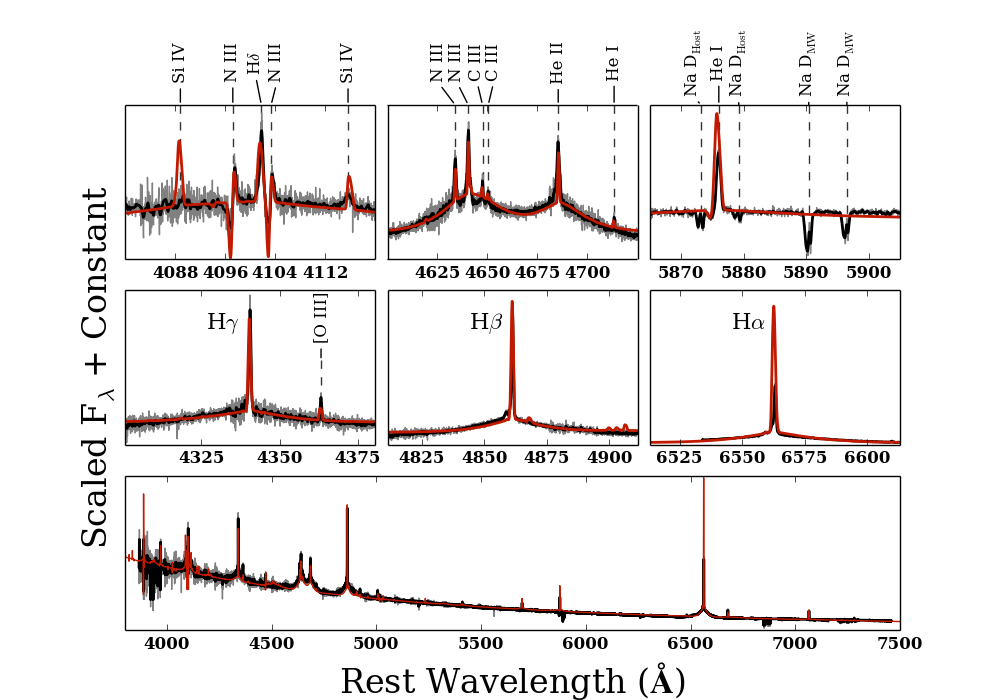}
\caption{Comparison between the observed spectrum of SN~1998S (black, smoothed to 0.5\,\AA~resolution) with the best-fitting CMFGEN model (red).
               The full-resolution spectrum is shown in the background (gray).
               Upper panels show individual lines of interest, while the lower panel illustrates the outstanding agreement
               over the full wavelength range.  Note that arbitrary vertical shifts are included in the model in the upper panels to match the
               observed local continuum, for clarity,
               and (as described in the text) that the degree of host-galaxy dust absorption was fit as a free parameter.
              \label{fig:model_fits} }
\end{centering}
\end{figure}

\begin{figure}[h]
\begin{centering}
\includegraphics[width=0.495\textwidth]{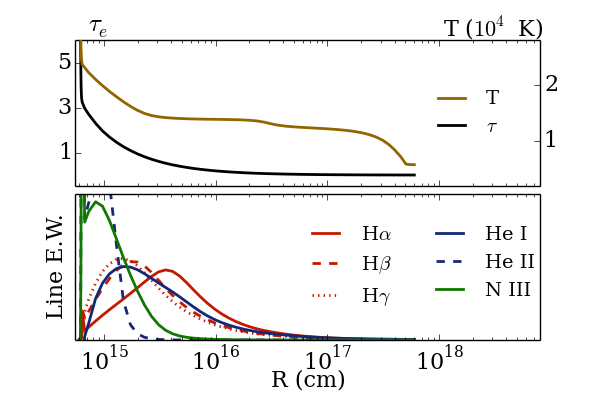}
\includegraphics[width=0.495\textwidth]{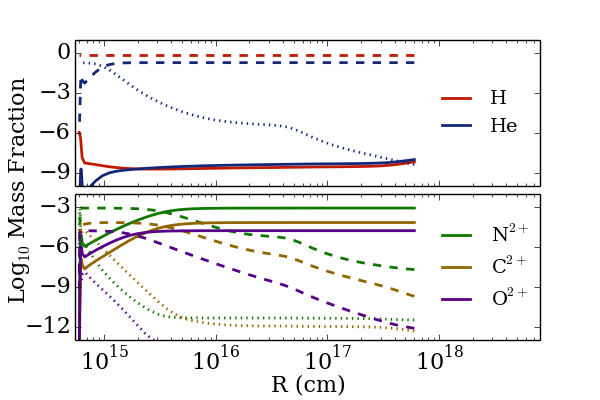}
\caption{Radial profiles of our best-fit CMFGEN model. The top-left panel
         shows electron optical depth and temperature, the bottom left
         shows emission for several prominent lines, the top right shows
         the ionization structure of H and He, and 
         and the bottom right shows the ionization structure of carbon, 
         nitrogen, and oxygen.  In the two right-hand panels, dashed lines
         indicate one ionization state higher and dotted lines
         indicate two ionization states higher. 
         In all profiles, the edge of the cool dense shell is apparent near
         $6 \times 10^{14}$\,cm, with the CSM continuing out
         to very large radii. \label{fig:model_profile}}
\end{centering}
\end{figure}

Figure~\ref{fig:model_fits} shows the model compared to the observed spectrum of SN~1998S, Table~\ref{tab:model} lists the parameters of our best-fit CMFGEN model, and Figure~\ref{fig:model_profile} plots the radial profiles of several model parameters.  The CMFGEN model does an excellent job and qualitatively matches the observed spectrum, predicting the appearance of almost all observed lines with the correct morphology. The quantitative agreement is relatively good, although some discrepancies are apparent. For each species, we endeavored to fit our model to all identified lines in an averaged sense; in practice, of course, this produces slight overestimates for some lines and underestimates for others.  
We include these uncertainties in our error estimates for model parameters --- the best-fit model described here is a compromise taking into account the various diagnostics we have, and small changes in the input parameters (within the error range given) may provide a better fit to individual spectral lines but the overall fit would deteriorate.
This can be seen in Figure~\ref{fig:model_fits}.  For example, our model fits H$\gamma$ well, though the electron-scattering wings of H$\beta$ are underestimated and the narrow component of H$\alpha$ is overestimated (discussed in more detail below), while for \ion{He}{1} our model overestimates \ion{He}{1}\,$\lambda$5875 and very slightly underestimates \ion{He}{1}\,$\lambda$4713.
Note that, to obtain a better fit, we have allowed the total amount of dust reddening to vary, and we find that the models prefer a smaller amount of host-galaxy dust than discussed in \S\ref{sec:obs} --- see \S\ref{sec:reddening}.

\begin{table}[H]
\centering
\begin{tabular}{l c c c}
\hline
\multicolumn{4}{c}{CMFGEN Model Parameters} \\
\hline
\multicolumn{2}{r}{ $L_\mathrm{SN}$ }  &  \multicolumn{2}{l}{ $(1.5 \pm 0.5) \times 10^{10}$\,L$_{\odot}$ }  \\
\multicolumn{2}{r}{ $R_\mathrm{in}$ }   &   \multicolumn{2}{l}{ $(6.0 \pm 1.0) \times 10^{14}$\,cm } \\
\multicolumn{2}{r}{ $\vwind$ } &  \multicolumn{2}{l}{ $40 \pm 5$\,\kms }  \\
\multicolumn{2}{r}{ $\mdot$ } &  \multicolumn{2}{l}{ $(6.0 \pm 1.0) \times 10^{-3}$\,M$_{\odot}\,yr^{-1}$ }  \\ 
\multicolumn{2}{r}{ $Y$ }  &  \multicolumn{2}{l}{ $0.49 \pm 0.10$ } \\
\multicolumn{2}{r}{ $E(B-V)$ }  &  \multicolumn{2}{l}{ $0.05$\,mag } \\
\hline
Element & Species & Mass Fraction ($\chi$) & $\chi/\chi_{\odot}$ \\
\hline
H & I & 0.4953 & 0.70\\
He  & I-II & 0.4928 & 1.8 \\
C  & II-IV & $5.0 \times 10^{-4}$ & 0.16 \\
N   & II-V & $7.2 \times 10^{-3}$ & 6.5 \\
O  & II-VI &  $1.6 \times 10^{-4}$ & 0.017 \\
Si  & II,IV  & $7.0 \times 10^{-4}$ & 1.0 \\
P  & IV,V  & $6.1 \times 10^{-6}$ & 1.0 \\
S   & IV-VI  & $3.7 \times 10^{-4}$ & 1.0 \\
Fe  & III-VII & $1.4 \times 10^{-3}$ & 1.0 \\
\hline
\end{tabular}
\caption{The best-fit CMFGEN model parameters. Listed error estimates are obtained by varying the input model parameters
                 and comparing the result to our observed spectra, and they include variance in the overall continuum shape and
                 in the relative line intensities within a species. \label{tab:model}}
\end{table}

The value of $R_\mathrm{in}$ and  $L_\mathrm{SN}$ determines the temperature at the inner boundary ($T_\mathrm{in}$), which is the primary regulator of the ionization structure of the illuminated progenitor wind. Our main diagnostic for the ionization structure is the ratio of \ion{He}{1} and \ion{He}{2} lines, which are very sensitive to $R_\mathrm{in}$ and  $L_\mathrm{SN}$ in this parameter range. 
To minimize the confounding effect of the uncertainties in the flux calibration and dust reddening, we use neighboring lines when setting these parameters: \ion{He}{1}\,$\lambda\lambda$4471, 4713 and \ion{He}{2}\,$\lambda$4686, all of which are fit reasonably well by the final model. 
Our best-fit values for $R_\mathrm{in}$ and  $L_\mathrm{SN}$ are in extremely good agreement with predictions for this epoch by \citet[][see their Fig.~2]{chugai01}, which are based upon
light-curve model fits to bolometric flux estimates from March~14 and onward \citep{fassia00}.
Additional constraints on the ionization structure come from the absence of \ion{N}{4}\,$\lambda$4057 and \ion{N}{4}\,$\lambda$7123.  To fit the absolute-flux level, we assume a distance of $d=17.00$~Mpc \citep{tully98} and, as explained in \S\ref{sec:reddening}, a standard Milky Way dust-extinction law \citep[$R_V=3.1$,][]{cardelli89}, finding a best-fit total dust reddening of $A_\mathrm{V}=0.15$\,mag or $E(B-V)=0.05$\,mag.

Given the high spectral resolution, we are able to fully resolve both the narrow and broad components of the features. This allows us to independently determine  $\vwind$ and $\mdot$. We find a progenitor wind speed of  $\vwind=40 \pm 5$\,\kms. To constrain this value, we use the width of the narrow component of the emission lines and the P-Cygni absorption component of \ion{N}{3}\,$\lambda$4097 and \ion{He}{1}\,$\lambda$5876. We obtain a progenitor mass-loss rate of $\mdot= (6 \pm 1) \times10^{-3}$\,\msunyr,  using the strength of H$\gamma$, H$\beta$, and \ion{He}{2}\,$\lambda$4686 lines as diagnostics.
 The advantage of using these lines rather than the stronger H$\alpha$ line is that they are formed in the inner parts of the outflow (Fig.~\ref{fig:model_profile}) and thus are less affected by time-dependent effects such as a variable progenitor $\mdot$, the changing radiation field from the SN, or light-travel time. Our model overestimates H$\alpha$, which may indicate either that part of the extended region responsible for the H$\alpha$ emission (Fig. \ref{fig:model_profile}) has not yet been illuminated by the SN, or that the density structure is steeper than $r^{-2}$, or both. 
Our best-fit value for \mdot~is within the range of previously-published estimates for the mass loss rate from SN~1998S's progenitor
\citep[$\mdot \approx 10^{-4}$--$10^{-2}$\,\msunyr, e.g.,][]{lentz01,chugai01,pooley02,moriya14}.

Figure~\ref{fig:model_profile} indicates that (except for the H$\alpha$ line) material between approximately $6 \times 10^{14}$\,cm and $2 \times 10^{15}$\,cm dominates the CMFGEN spectrum.
Assuming a constant \vwind~of 40\,\kms, this corresponds to material launched from the progenitor $\sim 5$--15\,yr before core collapse.
As shown in Table~\ref{tab:model}, our model indicates that the progenitor wind presents material mildly enriched by the CNO cycle, with He and N being enhanced and H, C, and O being depleted. The He mass fraction $Y$ is relatively well constrained since the spectrum of SN~1998S has strong H, \ion{He}{1}, and \ion{He}{2} lines. We use the relative strengths of the He and H lines as diagnostics and find $Y=0.49 \pm 0.1$. The N abundance is constrained using \ion{N}{3}\,$\lambda\lambda$4634, 4640 and the O abundance is determined from [\ion{O}{3}]\,$\lambda\lambda$4363, 4959, 5007.  We employ \ion{C}{3}\,$\lambda\lambda$4647, 4750, 5696 to constrain the C abundance, but the absolute value there should be taken with care given the well-known dependence of these C lines on details of the model atoms \citep{martins12}. We estimate uncertainties of 50\%, 30\%, and 50\% on the C, N, and O abundances, respectively. We assume solar abundances for the other elements, noting that several \ion{Fe}{4} lines that are predicted by the model are not observed. The absence of these lines could be reconciled with the observations either by decreasing $R_\mathrm{in}$ (i.e. increasing $T_\mathrm{in}$) or by decreasing the Fe abundance to 1/3--1/2 of the solar value.

\section{Discussion}
\label{sec:discussion}

\subsection{A Cartoon}
\label{sec:cartoon}

\begin{figure}[H]
\includegraphics[width=\textwidth]{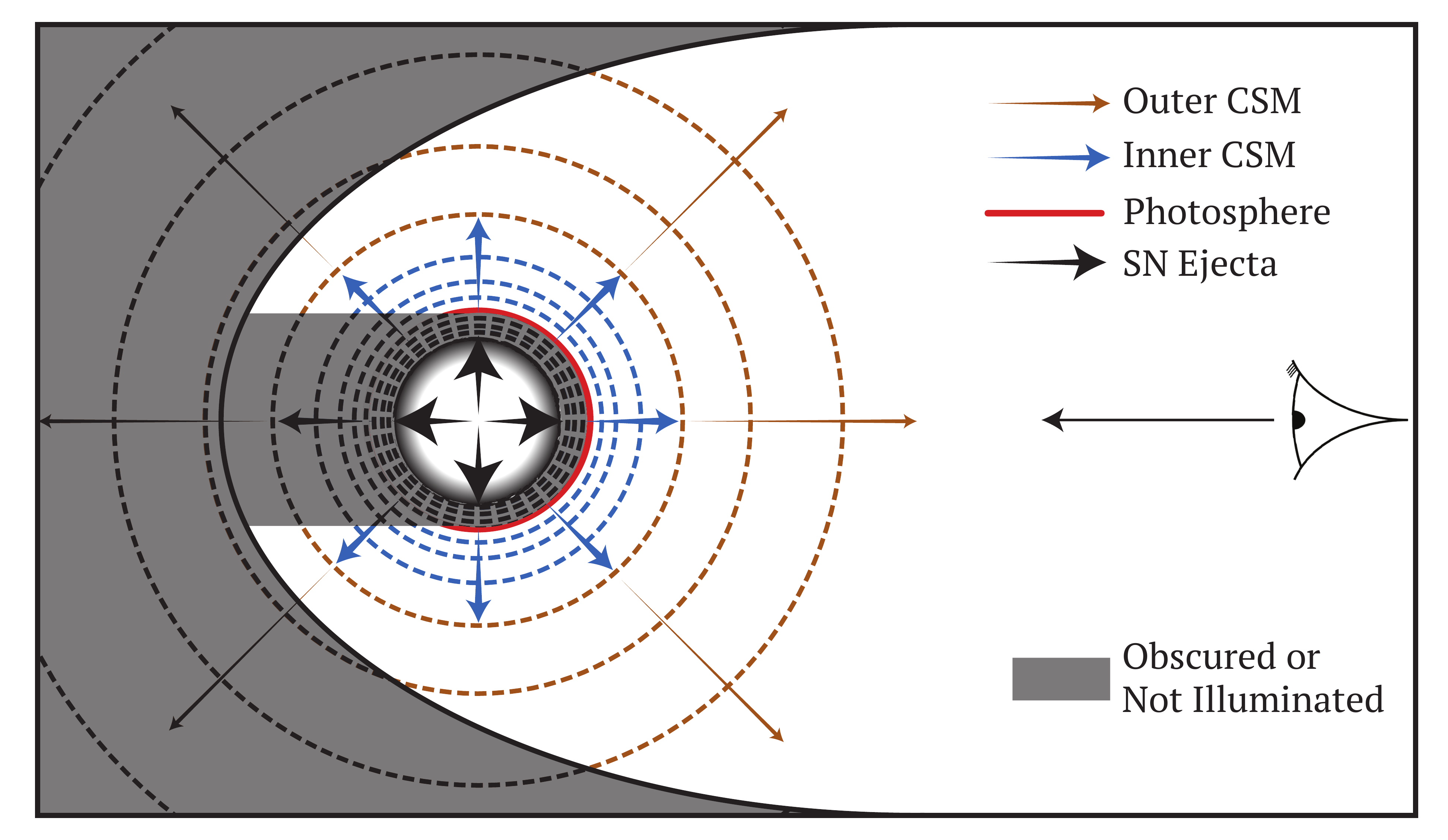}
\caption{A cartoon representation of SN~1998S and its circumstellar environment at the time of observation.  For the purposes
               of  this discussion, we define three zones:
               the SN ejecta (shown in black), the inner CSM region (blue), and the outer CSM region (orange).
               The photosphere of the system, which falls in the CSM, is shown in red.
               A fraction of the CSM is obscured by the photosphere and some regions have not yet
               received any SN radiation owing to nonzero light-travel time;
               these effects are illustrated by the gray mask, assuming the observer is to the right  
               (figure not to scale). 
               \label{fig:cartoon}}
\end{figure}

Figure \ref{fig:cartoon} displays a simple cartoon of SN~1998S and its CSM at the time of our HIRES observation.
SN~1998S was a hydrogen-rich core-collapse SN, and the expanding SN ejecta exhibited
radial velocities of $v \approx 7000$\,\kms\ \citep{leonard00}.
As indicated by our CMFGEN modeling (\S\ref{sec:98s_model}), the circumstellar material surrounding SN~1998S was quite dense. On day 2, when this
HIRES spectrum was taken, the photosphere was outside the expanding SN ejecta and within the extended progenitor wind.
In Figure \ref{fig:cartoon} the photosphere is drawn in red, the SN ejecta and swept-up CSM is shown at the center with black arrows,
the inner CSM is shown in blue, and the outer CSM is shown in orange.
This scenario is informed and supported by several previous studies of SN~1998S
in addition to our CMFGEN model \citep[e.g.,][]{leonard00,chugai01,fassia01,mauerhan12}.
An Earth-bound observer is shown to the right.  The regions of the system that are either obscured by
the photosphere or are not yet illuminated to the observer owing to 
the nonzero travel time of light have been masked in gray.

Note that Figure \ref{fig:cartoon} is not drawn to proportion.  To orient the reader, here are rough scales for the system
at this time, based upon the CMFGEN model outlined in \S\ref{sec:98s_model}: the outer radius of the expanding ejecta is $\sim 6 \times 10^{14}$\,cm,
the radius at which $\tau_e = 1.0$ is $\sim 2 \times 10^{15}$\,cm, and SN radiation 
has diffused through the optically thick CSM region and traveled $\sim 5 \times 10^{15}$\,cm into the far hemisphere of the wind,
while the entirety of the near hemisphere has been illuminated.

To understand the effect of nonzero light-travel time on the observed spectrum, consider the system immediately after the shock
traverses the progenitor, as SN radiation is diffusing through the optically thick CSM and ionizing it.
Radiative recombination emission from the CSM along the line of sight between SN and observer
will follow the first SN light immediately, but an Earth-bound observer will not begin to 
detect recombination emission from material along other lines of sight until the ionizing radiation has reached that
material and the recombination emission has then had time to propagate back to the observer.  
For a system like SN~1998S, which has a significant amount of material extending to at least $10^{17}$--$10^{18}$\,cm 
\citep[$\sim 10$--100 light days;][]{mauerhan12}, this effect
may be important when interpreting early-time observations.  For example, CSM located at a radius of $\sim 2 \times 10^{15}$\,cm on the far
side of the SN (but not obscured by the photosphere) would contribute flux to the observed spectrum at a time lag of about 36\,hr
relative to CSM located at the same radius on the near side of the SN.

To further confound the issue, the density profile
(and therefore ionization potential) is a strong function of radius, and the ionizing radiation field is evolving rapidly
as the ejecta/CSM boundary moves outward and the shock interaction fades over time.  
Any observed spectrum will therefore include emission from material ionized at different times by different
radiation fields, emitting material will continue to produce recombination radiation for 
different timescales after ionization, 
and this material will be located at a diversity of radii and Doppler shifts relative to the observer.
These effects are usually negligible when considering SN spectra, but because of the very young age and high resolution of this spectrum
these complications are worth keeping in mind,
especially for the H$\alpha$ line emission, which arises at relatively large radii (see Fig.~\ref{fig:model_profile}).

This cartoon assumes (as does our CMFGEN model) that the progenitor system had a more-or-less spherically
symmetric CSM surrounding it, but evidence to the contrary was presented by \citet{leonard00} based upon 
spectropolarimetry from 1998 March 7.
\citet{wang01} also present spectropolarimetry obtained March 30 and later.
The interpretation of the spectropolarimetric data is obfuscated by the
uncertain degree of interstellar polarization within the host galaxy, and \citet{leonard00} present three distinct 
physical scenarios consistent with the data.  In agreement with \citet{chugai01}, we prefer
the second interpretation of \citet{leonard00}: the narrow emission features (dominated by unscattered recombination emission)
are unpolarized, and the continuum and broad emission wings are both polarized at the few-percent level.

This can be understood in a scenario with a CSM that exhibits a nonspherical but axisymmetric geometry with the axis of symmetry
roughly orthogonal to our line of sight.  This CSM configuration could be the result of axisymmetrically-enhanced mass loss from the progenitor star
\citep[e.g.,][]{bjorkman93,ignace96}, an effect which is regularly observed in massive stars \citep[e.g.,][]{coyne82,zickgraf85,schulte92}, or perhaps the result 
of gravitational interaction with a binary companion \citep[e.g.,][]{smith11b}.
In such a CSM the mean free path of an ionizing photon is greater where the density is lower, and
so the ionization front will be elongated in those directions, thereby imparting a significant polarization signal on the continuum and broad-wing flux.
The narrow (unscattered) lines are unaffected, regardless of the geometry.

Of course, our quantitative CMFGEN results are dependent upon the assumption of spherical symmetry, but a moderate axisymmetric CSM density enhancement
could impart a clear spectropolarimetric signal without affecting the qualitative results from our CMFGEN modeling,
and we emphasize that these early-time observations do not probe any possible asphericity of the SN ejecta itself but only the dense CSM.
It is difficult to address this issue further without detailed modeling and a more complete dataset, but here we summarize a few additional points:
(1) at early times the lack of broad ejecta lines implies that the covering fraction of optically thick CSM is $\sim 1$ (though it may still be aspherical);
(2) the CSM may exhibit different degrees of asphericity at
different radii (as would be expected from explosive mass loss from the progenitor, a scenario discussed later), which would mean that
a time series of observations is required to fully understand the system; and
(3) understanding the effects of asphericity is a persistent difficulty in our efforts to constrain and understand the mass-loss properties of massive stars.

\subsection{The SN Ejecta and Opaque CSM}
\label{sec:ejecta}

The broad P-Cygni features usually observed in SN spectra were not apparent in SN~1998S until
about 15\,d after discovery and they remained remarkably weak throughout the SN's photospheric phase \citep{leonard00,fassia01}.
The early absence of ejecta lines is easily explained: as we show in \S\ref{sec:98s_model}, the photosphere is exterior
to the expanding ejecta.  The presence of significant CSM also explains the weakness of the SN signatures, once they
do appear --- as \citet{branch00} explain, a significant amount of ``toplighting'' owing to the ongoing CSM/ejecta interaction
effectively mutes the SN lines relative to the continuum even after the photosphere has receded into the ejecta.
In addition, \citet{chugai01} describes the opaque (in the Paschen continuum) ``cool dense shell'' (CDS) between the 
outer shock boundary (where the ejecta are sweeping up CSM) and the inner shock boundary (where the 
reverse shock is propagating into the expanding ejecta).  He argues that this CDS stays optically thick
out to 40--50\,d and continues to obscure the SN ejecta --- a transition from an inner CSM with covering fraction $\sim 1$ to
an outer CSM with covering fraction $<1$ could explain the emergence of weak ejecta lines between days 15 and 50.

Regardless, \citet{leonard00} and \citet{fassia01} argue that the weak ejecta features,
once they do emerge from the continuum, are characteristic of a SN~II with H and He features alongside those from Fe, Si, and O.
Figure~\ref{fig:fullspec} shows the spectral evolution of SN~1998S alongside the CMFGEN model.
The velocity of the expanding ejecta is $\sim 7000$\,\kms\ as measured by the blue edge of the SN
features on day 25 \citep{leonard00}.

The time delay between core collapse and shock breakout can be significant in systems
with a dense CSM --- following \citet{chevalier11}, the diffusion time for the CSM surrounding SN~1998S
is $t_\mathrm{d} \approx 0.85$\,d.  This means that core collapse of the progenitor would have occurred almost a
full day before any radiation escaped the system.  If SN~1998S underwent core collapse 5\,d before this HIRES spectrum was taken,
the date of the latest nondetection from \citet{li98}, and assuming $v_{\rm ejecta}$ is roughly constant, 
the outer edge of the expanding ejecta would be at $\sim 3 \times 10^{14}$\,cm.  However, note that the ejecta are slowed significantly
as they sweep up the CSM, which implies that the early expansion velocity would be faster than 7000\,\kms\ and
therefore that the true radius of the outer ejecta edge is likely a bit larger --- in better agreement with
our modeled value of $\sim 6 \times 10^{14}$\,cm.

\subsection{The Inner CSM}
\label{sec:innerwind}

As mentioned in \S\ref{sec:analysis}, and as other authors have noted \citep{leonard00,anupama01,fassia01},
the strong H emission lines observed in SN~1998S's early-time spectra are well represented by the
sum of a narrow Gaussian profile and a broad modified Lorentzian profile.  \citet{chugai01} has shown
that these line profiles are created in regions with $\tau$ of a few, where the narrow recombination
emission lines are ``diffused'' through multiple electron-scattering events.  
The end result is a narrow line core (with a Doppler width indicative of the emitting material's velocity)
with broad, symmetric wings (which provide a measure of the optical depth in the emitting region). 
This electron-scattering process has been observed to produce similar broad, symmetric line profiles in several SNe~IIn
\citep[e.g., SNe 2010jl, 2011ht;][]{fransson13,mauerhan13b}, and our CMFGEN model reinforces this scenario in SN~1998S.

\begin{figure}[H]
\begin{centering}
\includegraphics[width=0.75\textwidth]{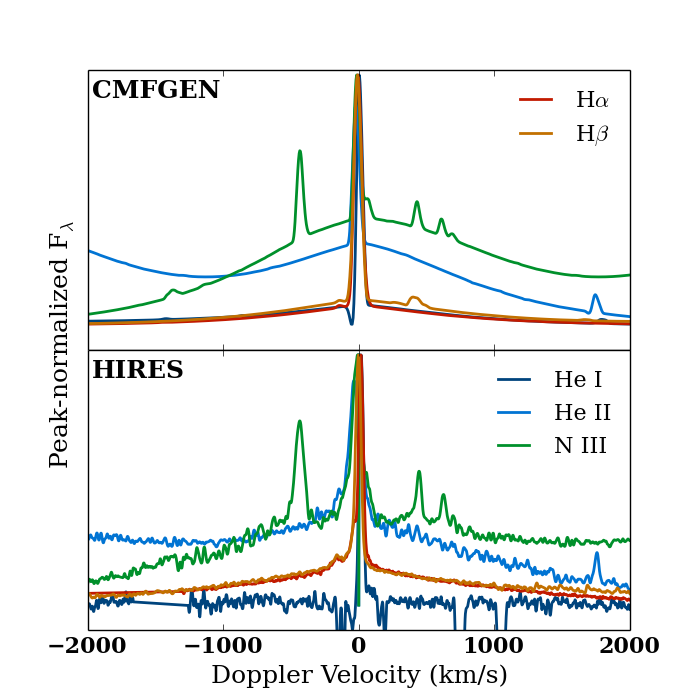}
\caption{A few notable line profiles normalized to their peak emission, shown for the CMFGEN model (top) and in the
              HIRES data (bottom, smoothed to a resolution of 0.5\,\AA).  We display H$\alpha$ (red), H$\beta$ (orange), \ion{He}{1} $\lambda$5876 (dark blue), 
              \ion{He}{2} $\lambda$4686 (light blue), and \ion{N}{3} $\lambda$4640 (green). 
              Lines that arise in the inner region exhibit more broad emission than lines that arise
              farther out in the CSM.  Note that the \ion{N}{3} complex here likely includes overlapping flux from the
              other nearby \ion{N}{3} and \ion{C}{3} lines.  In addition, any weak broad emission associated with the \ion{He}{1}
              line was likely below the noise threshold for detection in our raw spectrum --- if it exhibited a broad/narrow ratio comparable to that
              of the H$\alpha$ broad emission (as the CMFGEN model implies), it would have been indistinguishable from the 
              echelle order-response function and therefore subtracted out by our flux-normalization routine (see \S\ref{sec:obs}).
                         \label{fig:broad_narrow} }
\end{centering}
\end{figure}

The model further explains why some emission lines in the early SN spectrum exhibit these broad electron-scattering wings while others do not. 
Figure \ref{fig:model_profile} shows the best-fit model's radial profiles of the line-formation regions for
H$\alpha$, H$\beta$, H$\gamma$, \ion{He}{2}\,$\lambda$4686,  \ion{He}{1}\,$\lambda$5876, and \ion{N}{3}\,$\lambda$4640.
The model indicates that H$\alpha$ is formed over an extended region of the illuminated progenitor wind, with the bulk of the emission coming from $10^{15}$
to $2\times10^{16}$~cm ($1.6 \gtrsim \tau_e \gtrsim 0.06$), while the higher-ionization lines like \ion{He}{2}\,$\lambda$4686 and \ion{N}{3}\,$\lambda$4640
are formed in the inner region of the progenitor wind, from $6 \times 10^{14}$ to (2--3)$ \times 10^{15}$\,cm
($3 \gtrsim \tau_e \gtrsim 0.8$).
Figure~\ref{fig:broad_narrow} displays the line profiles of several of these lines normalized to their peak intensity, clearly showing that
emission lines which form primarily in the inner wind region (where there is a large electron-scattering optical depth) have strong, broad wings,
while lines that form primarily at larger radii (lower $\tau_e$) have relatively weak or no broad wings.
A similar result has been found for the Type IIn SN~1994W \citep{dessart09}, to which we refer the interested reader for an extensive discussion
on the formation of electron-scattering wings and its computation by CMFGEN. 

\begin{figure}[H]
\begin{centering}
\includegraphics[width=0.65\textwidth]{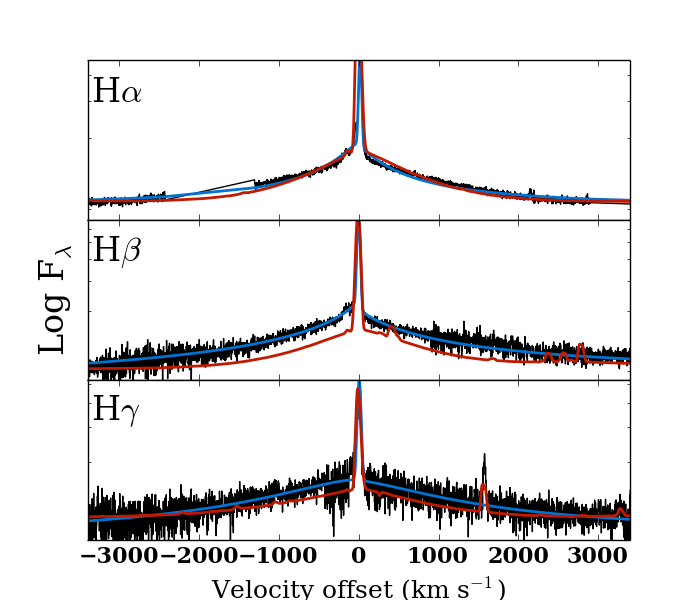}
\caption{A close-up view of the H emission lines on March 4 with the data in black, simple two-component fits
                shown in blue, and our CMFGEN model in red.
                All three H lines are well described by the sum of a narrow Gaussian and a broad modified Lorentzian,
                with the latter blueshifted by $\sim 50$\,\kms.  The CMFGEN model does not
                show this blueshift trend, as explained in the text.  Note that there is a weak helium emission line
                apparent $\sim 180$\,\kms\ blueward of both H$\alpha$ and H$\beta$, as well as other
                emission lines in the wings of all three lines.
                \label{fig:hlines}}
\end{centering}
\end{figure}

The HIRES spectrum presented here illustrates a further wrinkle: in all two-component lines,
the broad wings are systematically blueshifted from the narrow cores by $\sim 50$\,\kms.
Figure~\ref{fig:hlines} shows the first three lines of the Balmer series and their best-fit profiles, including
a narrow Gaussian profile, a broad modified Lorentzian profile, and a linear background.
Unfortunately, deconvolving all of the overlapping line profiles near the
\ion{N}{3}/\ion{C}{3}/\ion{He}{2} complex around 4650\,\AA\ is difficult and the simple
two-component fitting procedure we used for the H lines is not feasible for these ions, but a rough analysis indicates
that a similar offset between the central wavelengths of the broad wings and narrow cores exists.

We interpret these broad-wing blueshifts as evidence that the inner wind region exhibits higher
wind velocities than the outer region. The opaque photosphere preferentially obscures 
redshifted material more than blueshifted material, and the effect is more significant
for emission arising at small radii than at large radii --- 
as Figure~\ref{fig:model_profile} shows, the broad H emission (which depends upon the $\tau_e$ profile)
preferentially arises at lower radii than the narrow H emission.

\begin{figure}[H]
\includegraphics[width=\textwidth]{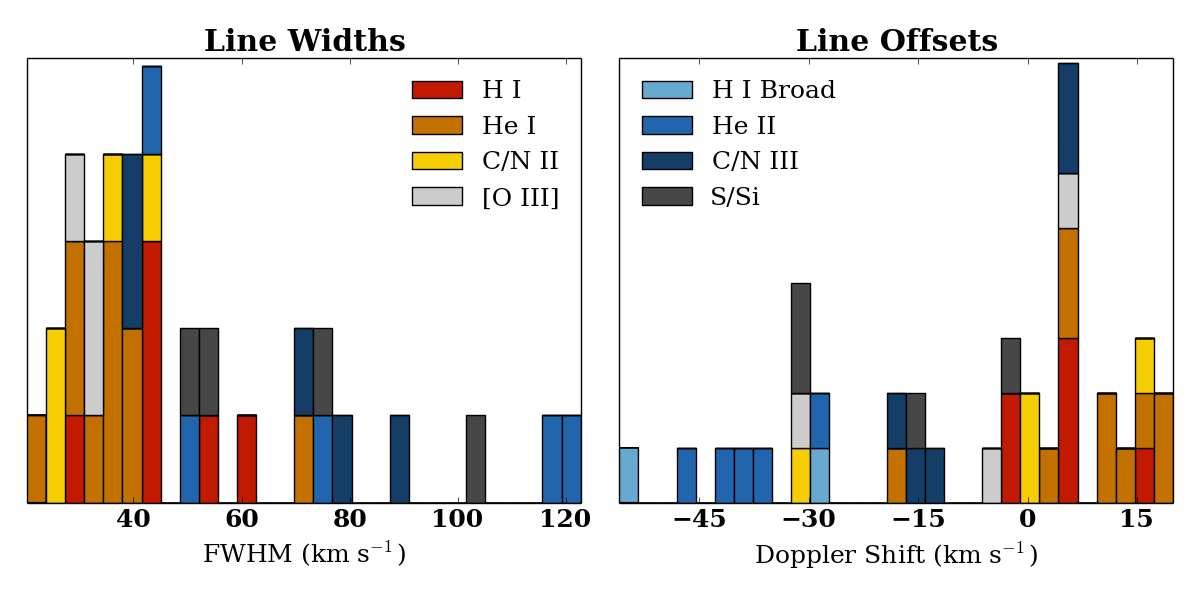}
\caption{A stacked histogram of line widths and line-center shifts for narrow emission lines in the HIRES spectrum of SN~1998S.
                Lines arising in the inner CSM are shown in red/orange/yellow, and lines arising predominantly in the outer CSM
                are shown in various shades of blue.  Inner CSM lines are, in general, broader and blueshifted from the outer CSM lines.
                A complete list of lines and their properties is in Table \ref{tab:lines}.  
                \label{fig:velocities}}
\end{figure}

Similarly, the narrow emission-line cores that arise at low radii are both broader
and more blue than the lines arising at large radii.
Figure \ref{fig:velocities} displays histograms of the observed line widths and line-center shifts for the emission
lines in Table \ref{tab:lines}.  Emission arising from higher-ionization states for C, N, and He display broader
line widths than their low-ionization counterparts do, as well as line centers that are significantly blueshifted
from those of the low-ionization lines.  This reiterates that the CSM at small radii, which exhibits
a higher ionization state and more significant obscuration from the photosphere,
is moving more rapidly than the CSM at large radii. 
Note that our CMFGEN model does not include hydrodynamical evolution and assumes a constant velocity profile in the CSM;
the observed trend in the line widths and centers is thus not captured in the model.

\begin{figure}[h]
\includegraphics[width=\textwidth]{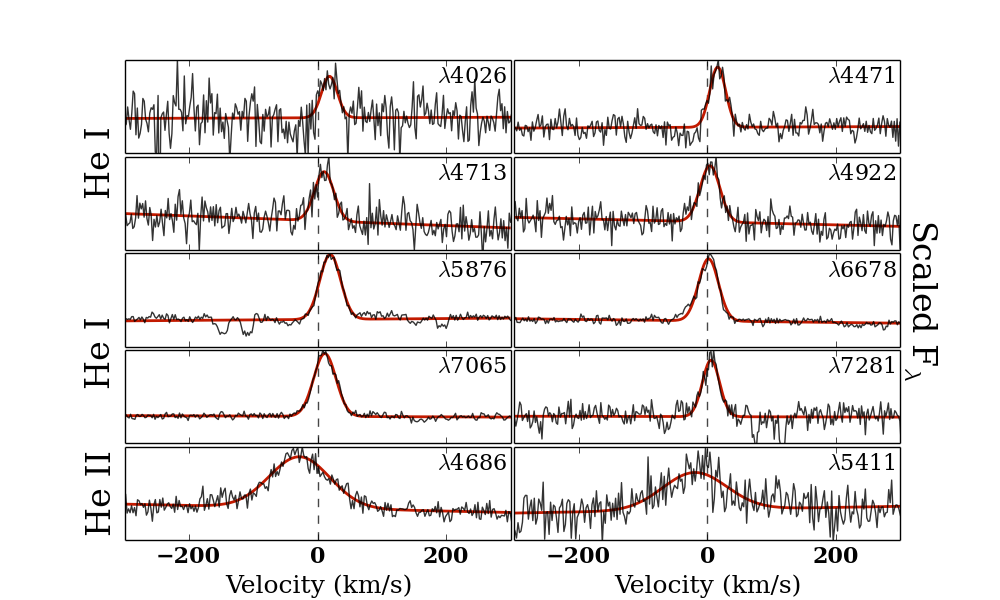}
\caption{Velocity structure of helium lines in the early HIRES spectrum of SN~1998S.
         \ion{He}{1} lines are shown in the top eight panels while \ion{He}{2} lines are in the bottom two.
         The HIRES spectrum is shown in black, red lines illustrate simple Gaussian fits to the data,
        and the vertical ashed lines indicate the rest frame.
               The \ion{He}{2} lines are significantly broader and blueshifted compared to the \ion{He}{1} lines.
               A complete list of lines and their properties is in Table \ref{tab:lines}.  
                \label{fig:helines}}
\end{figure}

The line-center effect is significantly smaller than the line-width effect, as expected,
and the \ion{He}{1} and \ion{He}{2} lines illustrate the effect nicely --- see Figure~\ref{fig:helines}.
We estimate the velocity of inner wind material by mirroring the blue wings of the \ion{He}{2} lines about 0\,\kms, 
which yields $v \approx 140$--180\,\kms, a factor of $\sim 4$ higher than the outer wind material.

Evidence for two distinct wind velocity components was also found in spectra 2+ weeks after explosion by \citet{bowen00} and \citet{fassia01},
with the same velocity for the slow component ($\sim 40$\,\kms) but with a significantly broader fast component ($\sim 350$\,\kms) ---
 \citet{fassia01} proposed that the faster wind was due to something like a blue supergiant phase
in the progenitor shortly before core collapse.  However, \citet{chugai02} presented models indicating a negative velocity gradient
in the CSM directly exterior to the photosphere, and proposed that the observations are better explained by the acceleration of a steady (slow) wind 
through radiation pressure from the SN rather than an increase in wind velocity shortly before core collapse.
This idea has been invoked to explain observations of other SNe~IIn as well \citep[e.g., SN~2010jl,][]{fransson13}.

Our observations are consistent with the latter scenario --- radiative acceleration naturally explains the observed
increase in inner wind velocity \citep[observations from March~20 indicate a velocity a factor of 2
greater than we observe on March~4,][]{fassia01}.
To explore this further, we adopt an approximation from \citet{chugai02}:
the maximum velocity to which the SN could accelerate a slow wind through Thomson scattering alone
is $v \approx 80\, E_{49} r_{15}^{-2}$\,\kms, where
$E_{49}$ is the time-integrated energy released through radiation in units of $10^{49}$\,erg and $r_{15}$ is the
radius in units of $10^{15}$\,cm.  We estimate $E_{49}$ by assuming that the luminosity of the young SN roughly
follows a power law in time: $L_{\rm SN} \approx \alpha (t-t_0)^n$.
We assume the time since explosion $(t-t_0)$ is $\sim 5$\,d and we determine $n \approx 3.25$ from early photometry
reported by \citet{li98} and \citet{qiu98}.  Adopting our modeled values for the luminosity and radius of the photosphere,
we calculate that $E_{49} \approx 0.6$ at the time of our HIRES spectrum.  This yields a maximum $v$ of $\sim130$\,\kms.
As \citet{chugai02} notes, including UV line-absorption effects increases the maximum velocity by $\sim50$\,\% in SN~1998S, 
and including the original wind velocity adds another $\sim 40$\,\kms.
This more than accounts for the full \ion{He}{2} velocity of $\sim 140$--180\,\kms\ measured from our HIRES spectrum.

In \S\ref{sec:cmfgen} we adopt a $\rho \propto r^{-2}$ wind profile for the CSM beyond the CDS.  This provides a 
good fit for most emission lines, but as Figure~\ref{fig:fullpagespec} demonstrates, our model overestimates
the narrow H$\alpha$ line.  The $\dot{M}$ parameter of the model adjusts the strengths of all hydrogen emission lines.
When fitting, we chose to match the H$\beta$, H$\gamma$, and H$\delta$ lines accurately, but
the H$\alpha$ line is better fit by a lower value.
As Fig.~\ref{fig:model_profile} shows, the H$\alpha$ line in the model includes flux at larger radii than the other
lines do --- our inability to match the entire Balmer series with a single $\dot{M}$ is evidence that the true
density profile in SN~1998S was steeper than $r^{-2}$.
As we discuss in more detail in \S\ref{sec:outerwind}, this is in good agreement with other
studies of SN~1998S \citep{chugai01,lentz01}.  Note, however, that a significant
fraction of the large-radii material is not yet illuminated owing to light-travel-time effects (Fig.~\ref{fig:cartoon}), and
this may also contribute to the observed effect.
 
Our best-fit value of the progenitor's $\dot{M}$ is $6 \times 10^{-3}$\,\msunyr.
This value, though remarkably high for most massive stars, is in good agreement with previous studies.
 \citet{chugai01} estimates the wind density parameter of SN~1998S through a combination
of the bolometric light curve and the line profiles, and \citet{moriya14} re-examine the bolometric light
curve with an updated model --- both estimate $\dot{M} \approx 10^{-2}$\,\msunyr.
To put this high $\dot{M}$ in perspective, the nearby red supergiant (RSG) Betelguese exhibits a steady mass-loss rate
of $(3-4) \times 10^{-6}$\,M$_{\odot}$\,yr$^{-1}$ \citep[e.g.,][]{glassgold86,harper01},
and luminous blue variables (LBVs) often have (variable) mass-loss rates a factor of
10--100 higher \citep{humphreys94}, as do yellow hypergiants \citep[YSGs;][]{lobel03}.
However, note that some rare RSGs can also exhibit extreme mass loss
 --- VY CMa is a nearby RSG undergoing variable and asymmetric mass
loss at a rate of $(2-4) \times 10^{-4}$\,\msunyr~at velocities up to 40\,\kms \citep{smith09}.

The radiative diffusion timescale and light-travel time across the inner CSM region are both less than 1\,d, and
the recombination times in this region are on the order of seconds to minutes
\citep[assuming the density and temperature profiles from \S\ref{sec:model};][]{osterbrock06,chevalier11}.
Thus, by the time our HIRES spectrum was taken, the inner CSM 
had already entered into photoionization equilibrium with the SN flux and
the flux coming from the ejecta/CSM shock boundary.

\begin{figure}[H]
\includegraphics[width=\textwidth]{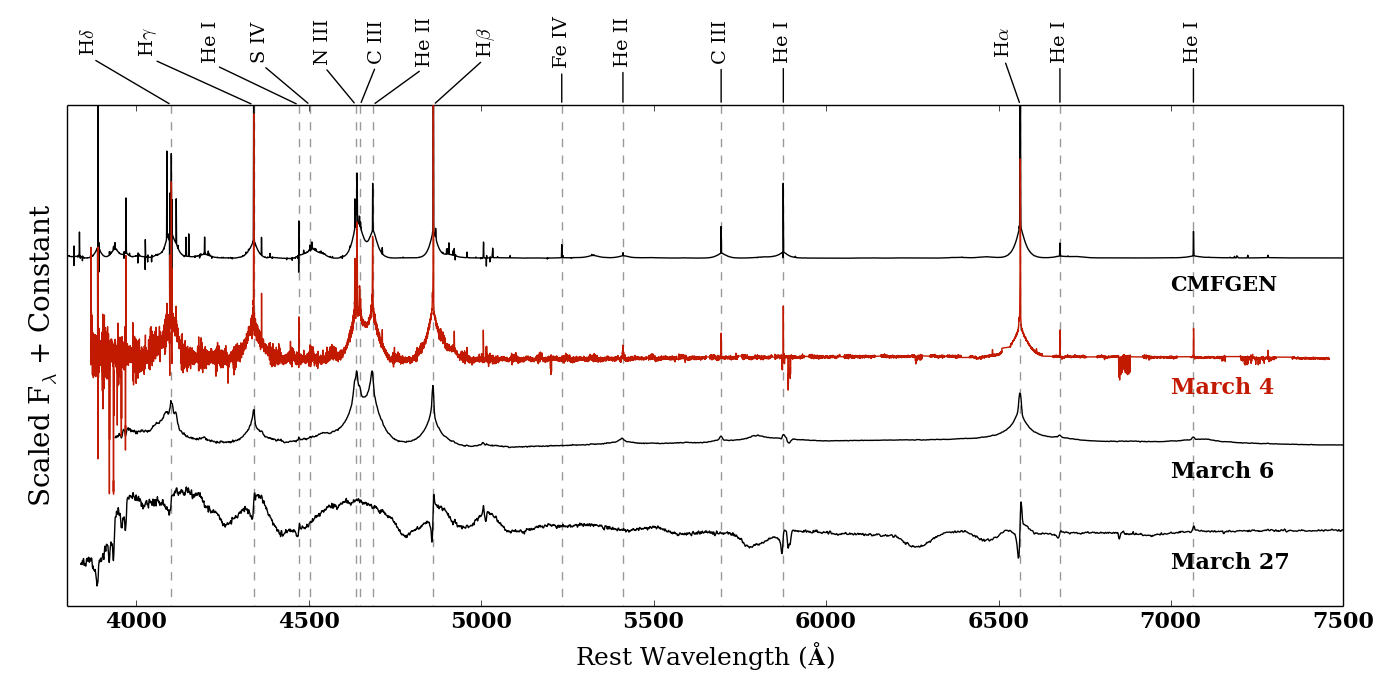}
\caption{The HIRES spectrum of SN~1998S and our CMFGEN model alongside low-resolution spectra
                taken at later phases \citep{leonard00}.
                A continuum fit has been subtracted from all spectra and notable features are marked with dotted lines.
                The transition from narrow emission lines to narrow P-Cygni profiles is apparent
                in the strong hydrogen and \ion{He}{1} lines, as all other narrow features fade away.
                The mismatch in iron abundance between model and data is apparent (\S\ref{sec:98s_model}) --- see the 
                \ion{Fe}{4}\,$\lambda$5234 feature.  In addition, it is probable that our flux-normalization procedure (\S\ref{sec:obs})
                removes some weak broad features --- e.g. see the \ion{Si}{4}\,$\lambda$4504, \ion{C}{3}\,$\lambda$5696, and
                \ion{He}{1}\,$\lambda$5876 lines.
                \label{fig:fullspec}}
\end{figure}

\subsection{The Outer CSM}
\label{sec:outerwind}

Several studies have established that the CSM around SN~1998S extends out to very large
radii \citep[e.g.,][]{gerardy00,leonard00,fassia01} and in observations taken $\sim 3$--14\,yr after core collapse
(when the expanding ejecta is at $r \gtrsim$ 6 $\times 10^{16}$\,cm) the H$\alpha$ emission indicates ongoing interaction
with a relatively vigorous RSG progenitor wind \citep[$\dot{M} \approx 10^{-4}\, \rm{M}_{\odot}$;][]{mauerhan12}.
Note that this mass-loss rate is $\sim 1.5$ orders of magnitude less than that inferred from our CMFGEN model.
\citet{lentz01} model UV/optical spectra of SN~1998S from March 16 (day 14), indicating that the CSM exterior
to $r_{\rm phot} \approx 10^{15}$\,cm exhibits an integrated $\tau$ of $\sim 0.2$.
This implies a steeper density gradient than $r^{-2}$, or a precipitous drop in wind density near $r \approx 10^{15}$\,cm
\citep[as proposed by][]{chugai01}, though as Figure~\ref{fig:fullspec} shows,
narrow P-Cygni profiles of H and \ion{He}{1} persist beyond March 25 --- strong evidence for at least some CSM at large radii.
Assuming the wind velocity remained constant at 40\,\kms, a density drop in the CSM at $r = 10^{15}$\,cm would correspond to a 
dramatic increase in the mass-loss rate off of SN~1998S's progenitor $\sim 10$\,yr before core collapse.
We note that the relative strengths of the Balmer emission lines in our early HIRES spectrum qualitatively support this scenario
(see \S\ref{sec:innerwind}) and our model is dominated by emission from material that left the star 5--15\,yr before
core collapse (see \S\ref{sec:98s_model}) --- a lower mass-loss rate 15+\,yr before core collapse is consistent with our model.

This timescale for increased mass loss is reminiscent of Galactic LBVs, which have been observed to undergo
extreme mass-loss events over timescales of 10--40\,yr \citep[e.g.,][]{humphreys94}, with one recent example
exhibiting an eruption only a few years before core collapse \citep[SN~2009ip;][]{mauerhan13,graham14}.
Alternatively, there could be a wind geometry transition near $r \approx 10^{15}$\,cm --- if the inner wind region were
roughly spherical while the outer region were in a disky or equatorial belt configuration,
the drop in CSM interaction luminosity could simply be caused by a drop in the CSM covering fraction.  
See \S\ref{sec:cartoon} for further discussion of asphericity in the SN~1998S system. 

As Figure \ref{fig:velocities} shows, the outer CSM emission lines in the HIRES spectrum
have a median Gaussian FWHM of $\sim 35$--40\,\kms, consistent with the velocities
of winds from extreme RSGs \citep[e.g.,][]{smith09}.  These narrow lines are fully resolved here,
which was not the case for the low-resolution early-time spectra previously published \citep[e.g.,][]{leonard00,fassia01}.

Unlike the inner CSM, which has had time to come into photoionization equilibrium, at this time the first radiation from the SN
was still propagating through the extended CSM region.
With a characteristic outer CSM recombination timescale of hours to days \citep{osterbrock06}, the observed spectrum must 
include emission arising from material ionized by the evolving radiation field at various time lags.
These effects are mitigated by the radial density profile in the CSM --- in an optically thin medium, emission roughly
traces the density of material, so CSM at
smaller radii (where light-travel-time effects are less important) should contribute most to the narrow-line emission in our spectrum.  
However, our model indicates that an appreciable amount of the Balmer and \ion{He}{1} line flux arises at $r > 10^{16}$\,cm, and
 \citet{mauerhan12} have shown that a significant amount of material is present even at $R \approx 2\times 10^{17}$\,cm.
At $\sim 5$\,d after core collapse (assuming a short diffusion time), only $\sim 5$\% of a sphere of material with that radius would be illuminated 
($\sim 84$\% for $r \approx 10^{16}$\,cm), so we must keep the light-travel effects in mind when considering the 
influence of material at large radii on the early-time spectra of SN~1998S.

Note that, as described in \S\ref{sec:obs}, our best-fit recession velocity is slightly less than that from \citet{fassia01}, which
was determined by fitting for the line centers of forbidden emission lines in a high-resolution spectrum from day 36.
Though the discrepancy is only $\sim 10$\,\kms, it is significant given the very high resolution data and the narrow line features.
\citet{fassia01} find that the P-Cygni peak of the H$\alpha$ line
moves from a redshift of $+8$ to $+17$\,\kms\ between days 17 and 36, while our spectrum on day 2 shows the peak of H$\alpha$ 
emission at a redshift of $+16$\,\kms\ (or $+26$\,\kms, adopting their value of $z$).
The cause of this line velocity evolution is unclear.

\subsection{Dust Extinction of SN~1998S}
\label{sec:reddening}

Both \citet{leonard00} and \citet{fassia01} have observed a blue ($\lambda \lesssim 5000$\,\AA)
flux excess in the very early spectra of SN~1998S, beyond what can be accounted for with
a blackbody continuum. Because the early-time spectrum of \citet{leonard00} provides the overall
flux calibration for the HIRES data presented here, the same issue remains.

However, the ionization states of lines in our detailed non-LTE CMFGEN model provide an
independent measure of the temperature, allowing us to compute a spectral energy distribution that can be compared to the observations.
We find remarkable agreement between the model and data if the total dust reddening toward SN~1998S is quite small: $E(B-V) \approx 0.05$\,mag,
assuming $R_V=3.1$.
Adopting $E(B-V) = 0.148^{+0.040}_{-0.028}$\,mag and the standard Milky Way dust-extinction law, as derived in \S\ref{sec:obs},
all CMFGEN models include a significant mismatch between the relative line fluxes and the continuum level in the blue region.
\citet{leonard00} explore reddened blackbody fits with $E(B-V)$ and $R_V$ as free variables
and find reasonable fits with total $0.08 \lesssim E(B-V) \lesssim 0.15$\,mag and $R_V \gtrsim 4.0$; our
results are in rough agreement, though we only vary $E(B-V)$ and hold $R_V$ constant.

Significant variance in the properties of dust extinction toward extragalactic
SNe has been observed before \cite[e.g.,][]{foley14}, likely owing to obscuration
from the CSM of the SN or from anomalous dust laws in the host galaxy.
Perhaps a similar situation presents itself here in SN~1998S --- there certainly is
a large amount of CSM in the immediate vicinity of the SN.  
Regardless, this uncertainty in the reddening only somewhat affects our determination of $R_\mathrm{in}$,  $L_\mathrm{SN}$, and $\mdot$ using CMFGEN, 
while values for $\vwind$ and the abundances are unaffected.
For instance, with $E(B-V) = 0.148$\,mag and $R_V = 3.1$, our model indicates $\mdot~\approx (7.5 \pm 1) \times 10^{-3} \,\msunyr$, $R_\mathrm{in} \approx 7.0\times10^{14}$\,cm,
and $L_\mathrm{SN} \approx 3\times10^{10}\,\lsun$.  These values are 17\%, 25\%, and 50\% different from the 
best-fit values when adopting the reduced value of $E(B-V) = 0.05$\,mag (see Table~\ref{tab:model}).
Note also that in \S\ref{sec:cartoon} we adopt an interpretation of the spectropolarimetric data from \citet{leonard00} that implies 
a relatively large amount of host-galaxy reddening based upon empirical relations between $E(B-V)$ and the degree of interstellar polarization from
Milky Way dust, though \citet{wang01} present an alternative interstellar dust polarization interpretation.
Adopting $E(B-V)_{\rm total} = 0.05$\,mag indicates that the polarimetric properties of the host galaxy's dust
likely differ significantly from those of the Milky Way.

\subsection{Wolf-Rayet, Luminous Blue Variable, Red Supergiant?}
\label{sec:wolfrayet}
 
As other authors have noted \citep[e.g.,][]{leonard00},
the low-resolution early emission-line spectrum of SN~1998S shows remarkable similarities to the spectra of Wolf-Rayet
(WR) stars.  However, as also found for SN~2013cu \citep{groh14} and PTF11iqb \citep{smith15}, this does not necessarily indicate that the progenitor of SN~1998S was a WR star.
We find that the emission lines observed in the early-time spectra of SN~1998S are a result of the photoionization of a pre-existing progenitor wind after the SN explosion,
and are physically unrelated to WR stars --- in good agreement with other analyses of SN~1998S and other SNe IIn.
Regardless, WR spectra generally fall within a well-accepted and physically understood classification scheme \citep[e.g.,][]{smith68,smith96} which can provide
a useful framework for comparison \citep[e.g.,][]{agy}.  \citet{groh14}, in an effort to clarify the distinction between WR spectra and young SN spectra,
suggests prepending an ``X'' to the WR spectral types when discussing spectra of post-explosion objects that exhibit WR-like spectra but are not WR stars.  
In this spirit, we apply the \citet{smith96} classification criteria to our spectrum of SN~1998S to explore the similarities and differences between this spectrum
of the young SN~1998S system and the spectra of WR stars.

First, the SN spectrum is nitrogen rich, similar to the WN class of WR stars.
After applying the flux and reddening corrections described in \S\ref{sec:obs},
the \ion{He}{2}\,$\lambda$5411\,/\,\ion{He}{1}\,$\lambda$5875 line ratio \citep[``ionization criterion'';][]{smith68}
and the relative strengths of the \ion{N}{3} and \ion{N}{4} lines
in our HIRES spectrum indicate similarities to the WN7 or WN8 spectral classes.
There is, of course, very strong hydrogen emission in the SN~1998S spectrum, with a ``hydrogen criterion'' \citep{smith96} greater than 1.3.
WR stars with a hydrogen criterion greater than 0.5 are given the ``h'' postfix, but
the value for our SN~1998S spectrum is well outside the range generally seen even in WNh stars.

Note that the WR classification scheme assumes that the major emission lines are fully resolved; for our HIRES spectrum 
that requirement is satisfied, but it may not be for lower-resolution spectra of SN~1998S or other similar objects.
Note, also, that both the ionization criterion and the hydrogen criterion are defined based upon the ratio of the narrow 
emission line peak to the local continuum and, when examined in detail, the line profiles in our HIRES spectrum display systematic
differences from those of WR stars.
For the ionization criterion this does not appear to be important, but if one compares integrated flux ratios rather than
peak flux ratios, thereby accounting for the broad wings of the 
H and the \ion{He}{1} lines, the hydrogen criterion is even larger than is given above --- highlighting again that the 
physical conditions in this post-explosion spectrum of SN~1998S are actually quite different from those found in WN stars.  
The ``strength-width'' criterion of \citet{smith96} is also not very informative when applied to the SN~1998S spectrum.

In summary, the early-time spectrum of SN~1998S is similar to WR spectra of spectral classes WN7h and WN8h, although
clear differences are apparent and they indicate a very different physical picture than in a WR star.
Following \citet{groh14}, our SN~1998S spectrum would be classified as an XWR7h or XWN8h, but
this classification scheme breaks down when one examines the details.

Our analysis of the early spectrum of SN~1998S provides three major clues to the nature of the progenitor --- 
its chemical composition ($Y=0.49$, N-enhanced material), mass-loss rate ($6 \times 10^{-3}\,\msunyr$), and wind velocity (40\,\kms).
Given the extreme mass-loss rate and the evidence for an increased mass loss soon before core collapse, the LBV interpretation becomes quite attractive,
although RSGs like VY~CMa \citep{smith09} and YHGs like $\rho$~Cas \citep{lobel03} also show evidence of outbursts with extreme and variable mass-loss rates.
In addition, our CMFGEN model indicates that the progenitor wind was only mildly enriched by the CNO cycle, while the spectra of Milky Way LBVs generally
indicate strong CNO enrichment \citep[e.g.,][]{davidson86}. 
Our physical interpretation is that the CSM around SN~1998S was created by mass loss from an SN progenitor that was
compositionally consistent with theoretical predictions of RSGs, YHGs, and LBVs at the pre-SN stage (\citealt{groh13a}; see \citealt{groh13b} and their Table 3).

Continued monitoring of SN~1998S over the years since explosion has not indicated any evidence for an older shell of very-high-density CSM,
as one might expect if there were previous outbursts.  However \citet{mauerhan12} publish spectroscopy of the H$\alpha$ line as late as $\sim 5000$\,d,
powered by interaction with a steady RSG-like progenitor wind at $r \approx 3 \times 10^{17}$\,cm --- material that (assuming $v_{\rm wind}$ is steady
at 40\,\kms) left the progenitor a few thousand years prior.  Of course, coverage is very spotty over those 5000\,d,
and increased interaction with a CSM shell caused by a short-lived high-mass-loss episode may have gone by unnoticed.
Variable and extreme mass loss in the few years before
core collapse was recently observed in detail in the SN~2009ip system \citep[e.g.,][]{mauerhan13,graham14};
perhaps the SN~1998S system underwent a similar process. Exactly how eruptive mass loss and the evolution toward core collapse
are (or are not) coupled in massive stars is poorly understood at best, and we look forward to further studies of this topic.

\subsection{SN 2013cu}
\label{2013cu}

In a remarkable spectrum of the Type IIb SN~2013cu taken less than a day after explosion, \citet{agy} discovered emission features
very much like those observed in our HIRES spectrum of SN~1998S.  The lower resolution of their SN~2013cu spectrum
makes the two-component emission features more difficult to analyze, but the H$\alpha$ profiles of the two SNe are almost
identical and most of the same emission lines are present in both (see their Extended Data, Figure 3). 
\citet{agy} interpret the SN~2013cu spectrum within the context of a WR-like wind and consider the possibility that 
it provides a long-sought observational connection between massive H-deficient WR stars and the
progenitors of stripped-envelope core-collapse SNe.

However, as shown in Figure 2 of \citet{agy},
most of the emission-line profiles in the early SN~2013cu spectrum are unresolved and
therefore must have widths $\lesssim 100$\,\kms, much slower than those expected for a WR wind.
The spectrum of SN~2013cu is similar to that of a WN6(h) star \citep{agy}, but
WN6 stars in the Milky Way have terminal wind velocities of $\sim 1800$\,\kms\ \citep{crowther07}.
The wings of the strongest few lines in the SN~2013cu spectrum do extend out to WR-like velocities 
if one interprets them as the result of a Doppler shift ($v \approx 2500$\,\kms). On the other hand, it is then
difficult to understand why the line cores (and the complete profiles of the weaker lines) are so remarkably narrow.

It instead seems likely that the broad wings of the emission lines in SN~2013cu were created by Thomson scattering
in an optically thick CSM moving at a low velocity, the same process that created the characteristic
emission-line profiles of SN~1998S.  \citet{agy} note this scenario as a possibility, though they adopt
$v_{\rm wind} \approx 2500$\,\kms\ in their analysis. 
\citet{groh14} presents a CMFGEN analysis of SN~2013cu, indicating that the broad-line profiles are indeed created by 
Thomson scattering, that $\vwind$ is low ( $\lesssim 100$\,\kms), and that the progenitor was likely an LBV or YHG.
\citet{agy} also note that the progenitor mass-loss rate indicated by their spectrum of SN~2013cu is very much
greater than the rates from known WR stars and models, which may be an indication that the mass-loss
rate increased markedly in the year or so before core collapse --- quite similar to the 
irregular mass-loss scenario we ascribe to the progenitor of SN~1998S.
The detailed emission-line profiles and the strong similarities between the early-time spectra of SN 1998S
and SN 2013cu cast further doubt on this particular connection between WR progenitors and stripped-envelope SNe.  
Rather than a persistent WR-like wind, it seems likely that a short-lived epoch of extreme
mass loss from the progenitor of SN~2013cu created the CSM whose flash-ionized lines populate the spectrum
seen in the first hours after explosion \citep[also suggested by, e.g.,][]{groh14}.

There is no evidence for an extended outer CSM region in SN~2013cu. The early-time emission features
fade quickly, unlike those of SN~1998S which are sustained by the ejecta/CSM shock front, and
the narrow P-Cygni H and He features seen in the spectra of SN~1998S do not appear.  Unlike SN~1998S,
SN~2013cu was hydrogen poor, exhibiting spectral features characteristic of a Type IIb SN.
The progenitors of these two SNe were
therefore quite different, with differing masses or, perhaps, with different evolutionary histories owing
to the effects of a nearby binary companion star, present in one progenitor system but not the other.
If both objects did indeed undergo extreme mass-ejection episodes in the 1--10\,yr before core collapse, it seems that
these episodes can occur in a variety of progenitor systems \citep[see also PTF11iqb,][]{smith15}.  The rapid ``flash spectroscopy'' technique described
by \citet{agy} will be a powerful tool to explore this issue further, enabling the study of the CSM immediately surrounding
SNe of various types before it is swept up by the expanding ejecta.

\section{Conclusion}
\label{sec:conclusion}

In this article we present what we believe to be the earliest high-resolution
($\Delta \lambda < 1.0$\,\AA) spectrum ever taken of a Type IIn supernova and the earliest published spectrum of SN~1998S:
a Keck HIRES spectrum taken only a few days after core collapse.
We present a CMFGEN model of the entire SN+CSM system and we explore the physical implications of this
remarkable spectrum in detail.
We discuss the connections between the progenitor of SN~1998S and massive stars in
the Milky Way Galaxy that are undergoing extreme mass loss and show that the progenitor of SN~1998S
shared many properties with Galactic examples of high-mass stars --- LBVs, YHGs, and some extreme RSGs.
We compare the spectrum of SN 1998S to that of the recent Type IIb SN~2013cu and find remarkable similarities,
indicating that the progenitors of qualitatively different SNe may exhibit similarly extreme mass loss 
shortly prior to explosion.

The spectrum published herein offers a new window into the complex circumstellar environment of a prototypical and well-studied
SN~IIn, and we hope it can be of some use as a case study as many more SNe are observed at very early epochs
in the coming era of autonomous spectroscopic studies, which will be capable of observing SNe within mere hours of discovery
\citep[e.g., the Spectral Energy Distribution Machine;][]{SEDM}.

\section*{Acknowledgments}

We thank Avishay Gal-Yam and Nathan Smith for useful discussions and comments,
and we are grateful to the organizers of the 2014 CAASTRO Annual Scientific 
Conference at which several helpful discussions of this work occurred.
We thank Tom Matheson for assistance with an early reduction
of the data.
The data presented herein were obtained at the W.~M.~Keck Observatory,
which is operated as a scientific partnership among the California
Institute of Technology, the University of California, and the
National Aeronautics and Space Administration (NASA); the observatory
was made possible by the generous financial support of the W.~M.~Keck
Foundation. We wish to recognize and acknowledge the very
significant cultural role and reverence that the summit of Mauna Kea
has always had within the indigenous Hawaiian community; we are most
fortunate to have the opportunity to conduct observations from this
mountain. This research has made use of the Keck Observatory Archive (KOA),
which is operated by the W. M. Keck Observatory and the NASA Exoplanet
Science Institute (NExScI), under contract with NASA.
This research has also made use of the NASA/IPAC
Extragalactic Database (NED) which is operated by the Jet Propulsion
Laboratory, California Institute of Technology, under contract with
NASA.

A.V.F.'s supernova group at UC Berkeley is supported through NSF grant
AST--1211916, the TABASGO Foundation, and the Christopher R. Redlich
Fund. Research by D.C.L. at San Diego State University is supported by
NSF grants AST--1009571 and AST--1210311.  This work was supported in
part by the NSF under grant No. PHYS--1066293. A.V.F. thanks the Aspen
Center for Physics for its hospitality during the ``Fast and Furious''
workshop in June 2014. J.H.G. is supported by an Ambizione Fellowship 
of the Swiss National Science Foundation.

{\it Facilities:} \facility{Keck:HIRES}

\bibliographystyle{apj}
\bibliography{bib}

\end{document}